\documentclass[lettersize,journal]{IEEEtran}
\usepackage{amsmath,amsfonts}
\usepackage{algorithm}
\usepackage{array}
\usepackage[caption=false,font=normalsize,labelfont=sf,textfont=sf]{subfig}
\usepackage{textcomp}
\usepackage{stfloats}
\usepackage{url}
\usepackage{verbatim}
\usepackage{graphicx}
\usepackage{cite}
\usepackage{algpseudocode}
\usepackage{algorithmicx}
\usepackage{multirow}
\usepackage{booktabs}
\begin{document}

\title{Balancing Information Perception with Yin-Yang: Agent-Based Information Neutrality Model for Recommendation Systems}

\author{Mengyan Wang, Yuxuan Hu, Shiqing Wu, Weihua Li, Quan Bai, and Verica Rupar

\thanks{Mengyan Wang, Auckland University of Technology, Auckland, New Zealand (e-mail: cjv2124@autuni.ac.nz).}
\thanks{Yuxuan Hu, University of Tasmania, Hobart, Australia (e-mail: yuxuan.hu@utas.edu.au).}
\thanks{Shiqing Wu, University of Technology Sydney, Sydney, Australia (e-mail: shiqing.wu@uts.edu.au).}
\thanks{Corresponding author. Weihua Li, Auckland University of Technology, Auckland, New Zealand (e-mail: weihua.li@aut.ac.nz). }
\thanks{Quan Bai, University of Tasmania, Hobart, Australia (e-mail: quan.bai@utas.edu.au).}
\thanks{Verica Rupar, Auckland University of Technology, Auckland, New Zealand (e-mail: verica.rupar@aut.ac.nz).}
}

\markboth{Journal of \LaTeX\ Class Files,~Vol.~14, No.~8, April~2024}%
{Shell \MakeLowercase{\textit{et al.}}: A Sample Article Using IEEEtran.cls for IEEE Journals}


\maketitle

\begin{abstract}
While preference-based recommendation algorithms effectively enhance user engagement by recommending personalized content, they often result in the creation of ``filter bubbles''. These bubbles restrict the range of information users interact with, inadvertently reinforcing their existing viewpoints. Previous research has focused on modifying these underlying algorithms to tackle this issue. Yet, approaches that maintain the integrity of the original algorithms remain largely unexplored. This paper introduces an Agent-based Information Neutrality model grounded in the Yin-Yang theory, namely, AbIN. This innovative approach targets the imbalance in information perception within existing recommendation systems. It is designed to integrate with these preference-based systems, ensuring the delivery of recommendations with neutral information. Our empirical evaluation of this model proved its efficacy, showcasing its capacity to expand information diversity while respecting user preferences. Consequently, AbIN emerges as an instrumental tool in mitigating the negative impact of filter bubbles on information consumption.
\end{abstract}


\section{Introduction}
\IEEEPARstart{O}{nline} social networks often emphasize content shaped by users' past behaviors. While user preference-based recommendation systems enhance user engagement, they might also trap users in ``filter bubbles'' \cite{michiels2022filter}. These bubbles can set users in their existing views, leading to an imbalanced understanding of certain topics.

For instance, a user is deeply fascinated with the health-centric diet, believing in its numerous positive impacts. Preference-based recommendation algorithms may continuously recommend this user with favorable content about the health-centric diet, seldom presenting its drawbacks. Over time, this can reinforce the user's positive stance, thus creating a biased perspective and encapsulating the user within a filter bubble. This phenomenon isn't exclusive to the topic of health. Filter bubbles span various topics, sharing the common risk of fostering biased perceptions \cite{schumann2020we}. The societal impact of these bubbles is profound, which affects individual cognition and polarizes communities \cite{liu2023emotion,flaxman2016filter}. This emphasizes the importance of information neutrality, mitigating biases from filter bubbles by offering broader, balanced information.

Current solutions typically modify the underlying algorithms of recommendation systems to alleviate the filter bubble effect. However, such approaches often require substantial changes and potentially compromise user engagement and satisfaction. This paper integrates the Chinese Yin-Yang ($Y_-\&Y_+$) theory into recommendation systems, proposing an Agent-based Model for Information Neutrality (AbIN). This model aims to address imbalances in information perception and mitigate the effects of filter bubbles while preserving the functionality of existing recommendation algorithms.

We model the recommendation process of AbIN in a distributed manner utilizing Agent-based Modeling \cite{li2017agent}. Here, the original recommendation algorithms and the system users are conceptualized as two distinct agents: the Original Preference-based Agent (OPA) and the User Agent (UA). Furthermore, we embed a proactive Information Neutrality Agent (INA) to mediate the interactions between the OPA and UA, aiming to balance information perception and incorporate user feedback. This design ensures that the AbIN can neutralize information without changing the core recommendation algorithm.

The AbIN model draws inspiration from the $Y_-\&Y_+$ theory, emphasizing the balance between contrasting elements, thereby promoting a well-rounded understanding of diverse topics. The $Y_-\&Y_+$ theory, foundational in Chinese philosophy, is symbolized by a circle divided into two halves by an `S'-shaped curve. These two halves express the complex relationship between opposites (opposite sentiments on a specific topic). $Y_-$ represents negativity and darkness, while $Y_+$ means positivity and light \cite{kandel1998intrinsic}. Achieving $Y_-\&Y_+$ neutrality in recommendations is balancing these opposing sentiment energies, and providing users with broader choices rather than limiting them to specific preferences. In light of this philosophy, our model seeks to mitigate information sentiment imbalances, counter filter bubbles, and deliver a balanced recommendation experience to users. The contributions of the proposed approach are summarized as follows:

\begin{enumerate}
\item{Firstly, we are the first to invoke the Chinese $Y_-\&Y_+$ theory in recommendation systems. We propose a $Y_-\&Y_+$ Neutralization Control (YYNC) method to provide users with balanced information, broaden their exposure to diverse viewpoints, and foster a more holistic understanding of information.}
\item{Secondly, we adopt a distributed modeling approach to build a novel model, AbIN, aimed at mitigating the effects of filter bubbles. This model ensures the continued usability of existing recommendation systems without requiring any changes to their algorithms. The AbIN model incorporates multiple independent agents, i.e., OPA, UA, and INA. These agents can interact with each other. This agent-based model enhances the system's flexibility, allowing for the independent optimization of each agent. Moreover, it reduces implementation risks and provides a more effective solution to address the challenges posed by filter bubbles.}
\item{Thirdly, we carried out extensive experiments. The result establishes the AbIN model's efficacy in balancing recommended content, diminishing the impact of filter bubbles, and enhancing users' well-rounded understanding of the information they consume.}
\end{enumerate}

\section{Related Works}
\label{sec:related_works}
The increasing prominence of filter bubbles due to preference-based recommendation systems has gained scholarly attention. This section reviews existing literature focusing on the need for information balance in online platforms, the role of recommendation algorithms in forming filter bubbles, and methodologies for measuring the filter bubble effect.

To begin, online platforms can inadvertently fuel polarized viewpoints if there is an imbalance between positive and negative information. Fernandes \cite{fernandes2023confirmation} delves into the impact of confirmation information bias on social learning networks. The study introduces a theoretical model examining how agents exchange information influenced by their biases. On a similar note, Li et al. \cite{li2023double} investigate both the uplifting and detrimental effects of social networks. Different from these studies, which emphasise categorizing information as either positive or negative, our work more comprehensively considers information diversity and emotions based on the $Y_-\&Y_+$ theory.

Subsequently, AI-based recommendation systems have become increasingly sophisticated, tailoring content based on user behavior and preferences \cite{guo2020attentional}. Despite their convenience, they tend to perpetuate filter bubbles. Conventional algorithms like collaborative filtering \cite{wang2021robust} and content-based filtering \cite{reddy2019content} have been critically examined for this. Studies confirm that these recommendation methods can contribute to the formation of filter bubbles by limiting exposure to a variety of perspectives \cite{vilela2021majority,geschke2019triple,hu2022AI}. In contrast to existing works, our study extends this discussion by looking not just at content diversity but also at the sentiment behind users' consumption.

Lastly, attempts to mitigate the filter bubble effect are diverse. Lunardi et al. \cite{lunardi2020metric-news} recommend the k-nearest neighbors (k-NN) item-based recommendation approach with the Maximal Marginal Relevance (MMR) algorithm and conclude that the diversified recommendation strategy can reduce homogenization, as it offers users a wide range of topics. Su et al. \cite{su2020diversifying} address the low-diversity responses in open-domain dialogue generation by incorporating non-conversational text, which yields significantly more varied responses while maintaining contextual relevance. Grossetti et al. \cite{grossetti2021reducing} introduce a community-aware model to identify category-based similarities within Twitter communities, where a re-ranking model is integrated into the recommendation algorithm to effectively enhance the diversity of recommendation outcomes. Furthermore, researchers have explored graph-based approaches as a means to tackle this issue. For instance, Yang et al. \cite{yang2023dgrec} adopt graph-based techniques to enhance the diversity of recommendations, focusing on refining the embedding generation process. Similarly, Li et al. \cite{li2023breaking} optimize the generation of recommended lists within the user feedback loop by analyzing data in the user-item interaction graph. These approaches usually focus on modifying existing recommendation algorithms to counteract the filter bubble effect. It may not be applicable to systems that are already deployed, or it may affect user engagement. In contrast, our research presents a novel model to neutralize filter bubble influences without altering existing recommendation algorithms.

\section{Framework and Formal Definitions}
\label{sec:ffd}
\subsection{Overall Framework}

\begin{figure*}[!t]
\centering
\includegraphics[width=7.0in]{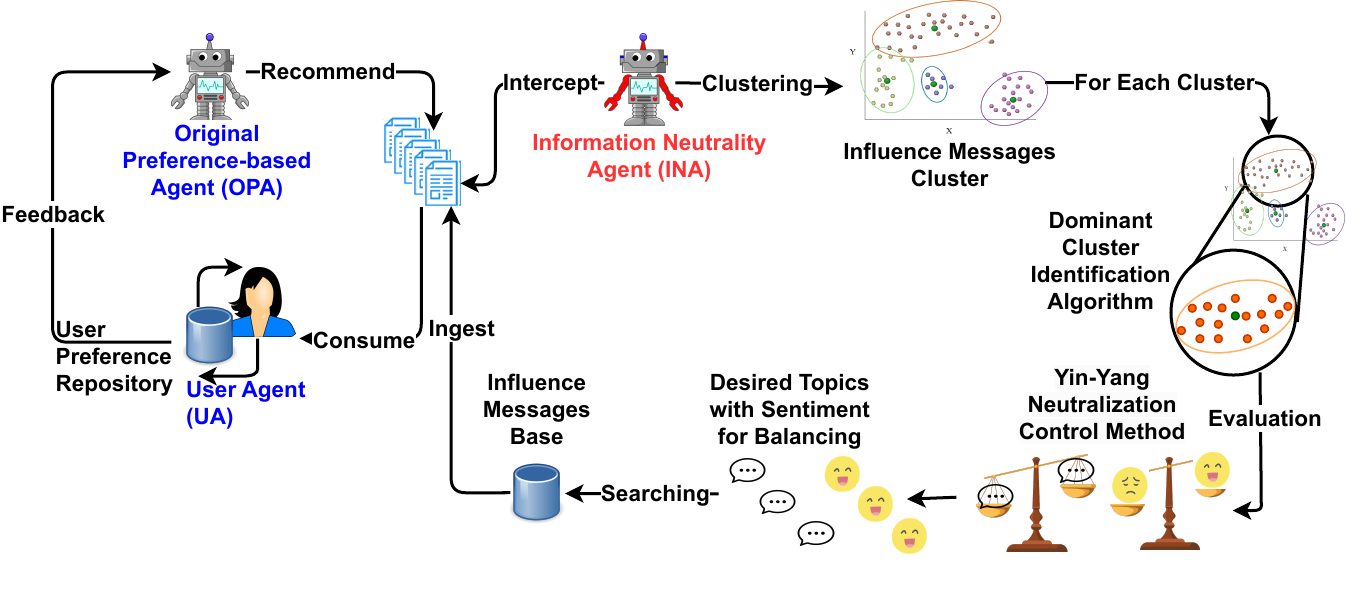}
\caption{The Framework of Agent-based Model for Information Neutrality (AbIN).}
\label{fig:framework}
\end{figure*}

In this research, we leverage agent-based modeling to develop the AbIN model, aimed at mitigating filter bubbles. As displayed in Figure \ref{fig:framework}, the AbIN model comprises three independent yet interconnected agents: OPA, UA, and INA. Each agent employs its unique learning strategies. A user is represented as an interactive User Agent (UA) with specific preferences. The Original Preference-based Agent (OPA) functions as the conventional preference-based recommendation system, learning UA's preferences and suggesting messages that align with these preferences. To counterbalance the potential bias introduced by OPA, we introduce the Information Neutralization Agent (INA). INA's role is to analyze and adjust the recommendation list from OPA by applying the Yin-Yang Neutralization Control (YYNC) method, ensuring a balance of viewpoints. After processing, INA forwards this balanced list of recommendations to UA. Through this mechanism, INA integrates complementary viewpoints into OPA's recommendations, thereby offering UA a balanced information experience. This approach aims to harmonize the viewpoints within the recommendations, ensuring diversified and unbiased information to the user.

\subsection{Formal Definitions}
In this subsection, we introduce formal definitions crucial for an understanding of the concepts central to AbIN.

\noindent \textbf{Definition 1: User Agent ($v_i$).} \label{eq:ua} This term refers to an individual within a social context, whose representation is captured through an embedding process. At any given time $t$, the embedding of $v_i$ ($\mathbf{v_i}$) is computed by aggregating the embeddings of all messages interacted with by $v_i$, utilizing the Hadamard product \cite{million2007hadamard} for combination. UA receives recommendations from INA and decides whether to accept or reject these recommendations based on their preferences.

\noindent \textbf{Definition 2: Influence Messages ($M^t_{v_i}$).} The OPA recommends a set of messages at time 
$t$ based on the preference of $v_i$, denoted as $M^t_{v_i}$ = \{$msg_1$,...,$msg_j$\}. Each message, $msg_x$, comprises a tuple: $(q(msg_x), T(msg_x), o(msg_x))$, containing the message's content text $q(msg_x)$, the associated topic $T(msg_x)$, and the sentiment intensity $o(msg_x) \in[0, 1]$. In this paper, the $Y_-\&Y_+$ neutralization is the sentiment balance for specific topics in $M^t_{v_i}$. Therefore, it is important to classify $Y_-$ and $Y_+$, which are introduced in Definition 3.

\noindent \textbf{Definition 3: Yin and Yang.} Messages are classified as $Y_-$ if $o(msg_x)$ ranges from 0 to 0.5, and as $Y_+$ if it exceeds 0.5. An $o(msg_x)$ value of precisely 0.5 denotes a neutral sentiment. The sentiment's intensity is considered more extreme as its value diverges further from 0.5, whether it falls under $Y_-$ or $Y_+$.

\begin{figure}[!t]
\centering
\includegraphics[width=3.0in]{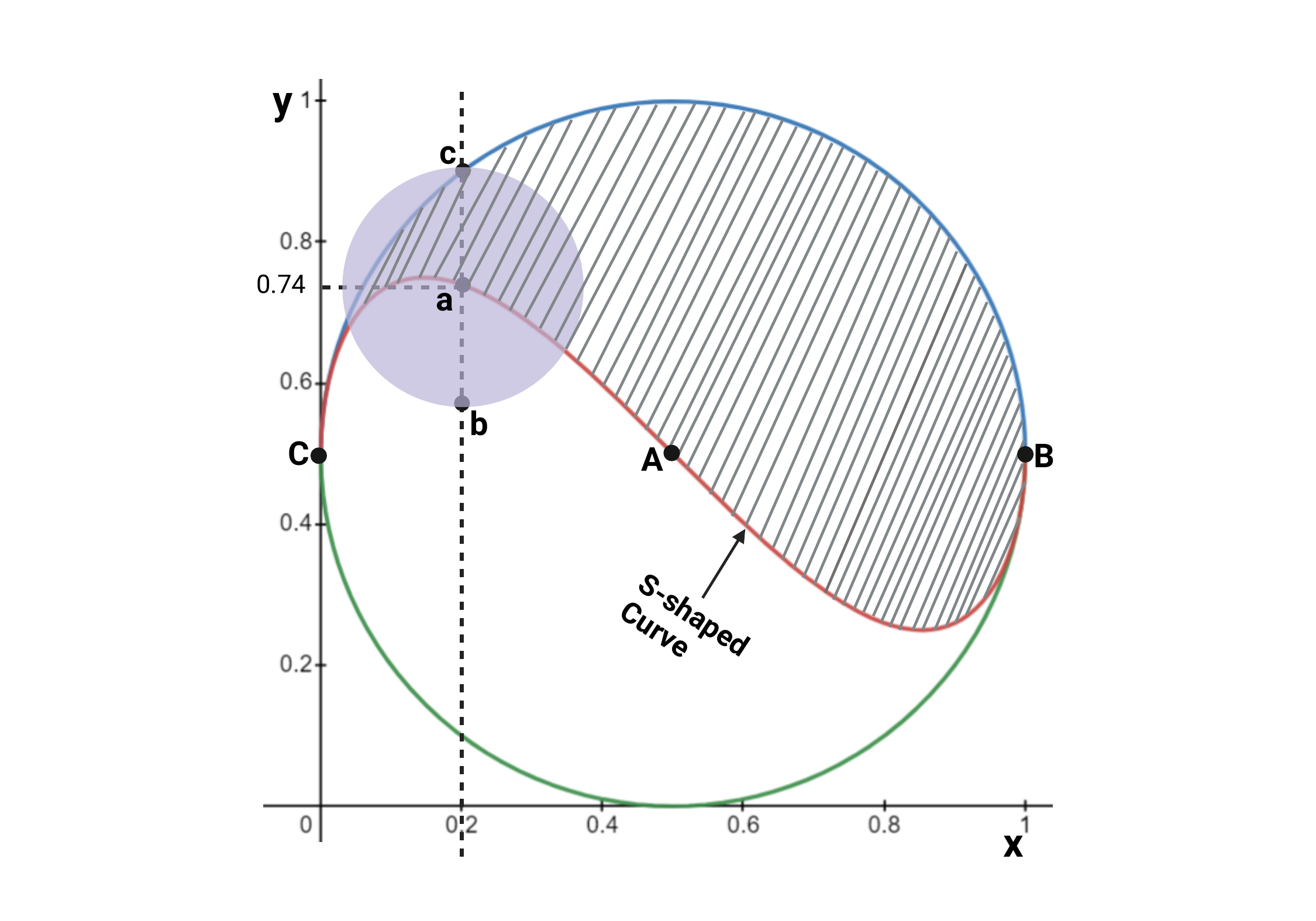}
\caption{The Chinese Yin-Yang ($Y_-\&Y_+$) Symbol.}
\label{fig:YINYANG}
\end{figure}

\subsection{Yin-Yang Neutralization Definition} \label{sec: YYND} Figure \ref{fig:YINYANG} depicts a conceptual diagram based on the Chinese $Y_-\&Y_+$ philosophy. It displays sentiment values ranging from 0 to 1 along both the x and y axes. This diagram features a circle centered at (0.5, 0.5) with a radius of 0.5, divided by an `S'-shaped curve (segment C-B) that represents the transition between $Y_-$ (above) and $Y_+$ (below) sentiments. This curve embodies the perfect $Y_-\&Y_+$ neutralization \cite{kandel1998intrinsic}. Inspired by \cite{chen2016mathematic}, the mathematical representation of the perfect $Y_-\&Y_+$ neutralization is given by:

\begin{equation}
\begin{aligned}
f(o^x_{\tau_{i}})=\frac{1}{2}+\left(1-2o^x_{\tau_{i}}\right)\sqrt{\frac{1}{4}-\left(o^x_{\tau_{i}}-\frac{1}{2}\right)^{2}}
\end{aligned}
\label{eq:stype}
\end{equation}

\noindent
where $o^x_{\tau_{i}}$ indicates a sentiment score in $O_{\tau_{i}}$ = \{$o^1_{\tau_{i}}$,...,$o^x_{\tau_{i}}$\}. $O_{\tau_{i}}$ includes sentiment scores for each message under topic $\tau_{i}$ in $M^t_{v_i}$. The point ($f(o^x_{\tau_{i}})$, $o^x_{\tau_{i}}$) is on the `S'-shaped curve and signifies a $Y_-\&Y_+$ neutralization pair for $o^x_{\tau_{i}}$ = $o(msg_x)$ within topic $\tau_{i}$.

The $Y_-\&Y_+$ neutralization aims for sentiment balance in $O_{\tau_{i}}$ for the specific topic $\tau_{i}$. The analysis process of balance using the following criteria: 1) We sequentially select a sentiment score $o^x_{\tau_{i}}$ from $O_{\tau_{i}}$ based on the sentiment extremity. 2) We establish a tolerance interval for $Y_-\&Y_+$ neutralization by drawing a circle centered at ($o^x_{\tau_{i}}$, $f(o^x_{\tau_{i}})$) with a radius equal to the shorter distance from x=$o^x_{\tau_{i}}$ to the curve's intersection. The rationale for establishing this interval is that users with extreme preferences tend to have a narrower acceptance threshold, whereas those with more moderate views possess a broader range of acceptance.

Referring to Figure \ref{fig:YINYANG}, when $o^x_{\tau_{i}}$ is 0.2, any sentiment score within $O_{\tau_{i}}$ that falls between points b and c can be considered a $Y_-\&Y_+$ neutralization pair with $o^x_{\tau_{i}}$, denoted as ($o^x_{\tau_{i}}$, $o^{x^+}_{\tau_{i}}$). Here, $o^{x^+}_{\tau_{i}}$ is seen as the complementary sentiment to $o^x_{\tau_{i}}$. If every value in $O_{\tau_{i}}$ can be paired in this way, then $O_{\tau_{i}}$ is considered to have achieved $Y_-\&Y_+$ neutralization.

\section{Agent-Based Information Neutrality Model}
\label{sec:strategies}
In this section, we detail the operations of the AbIN model, which encompasses three critical, interrelated yet independent agents: OPA, INA, and UA. In summary, OPA is tasked with creating messages based on user preferences and forwarding them to INA. INA intercepts these messages and outputs a final $Y_-\&Y_+$ neutralization recommendation list to UA. UA responds to outputs from INA based on current user preferences and deposits these responses into the user preference repository. Finally, OPA accepts responses from UA and provides the next set of recommendations to INA. 

\subsection{Original Preference-based Agent}

The Original Preference-based Agent (OPA) serves as the original preference-based recommendation system, employing a preference-based recommendation algorithm, like Collaborative Filtering (CF) \cite{wang2021robust} recommendation algorithm, to suggest a set of messages. OPA interacts with UA to select a list of messages based on the preferences generated by UA. As the input of INA, these selected messages are then processed for information neutralization through INA in preparation for generating the final recommendation list.

\subsection{Information Neutrality Agent}

The Information Neutrality Agent (INA) plays a crucial role in the AbIN model by facilitating information neutralization. Upon receiving a recommendation list from OPA, INA initially organizes the messages into clusters. Subsequently, we propose the Dominant Cluster Identification Algorithm (DCIA) to identify the target cluster for the $Y_-\&Y_+$ neutralization. Following this, we employ a novel Yin-Yang Neutralization Control (YYNC) method to achieve information balance within recommendations. INA then selects appropriate messages from the information messages base, guided by the $Y_-\&Y_+$ neutralization results from YYNC, and integrates these with OPA's recommendations. This enriched recommendation list is finally forwarded to UA.

\subsubsection{Message Clustering} \label{sec:msg_cluster}
Message clustering is an initial step in INA, setting the stage for subsequent information neutralization. This step involves grouping messages based on semantic and thematic similarities, which are quantified using both textual embeddings and topic representations. We employ BERT \cite{kenton2019bert} for generating textual embeddings for each message ($msg_x$). Each message is also tagged with a topic \(T(msg_x)\).

\textbf{Textual Similarity:} Cosine distance \cite{behrendt2021arguebert} between the embeddings of two messages, $msg_i$ and $msg_j$, is calculated as:

\begin{equation}
\begin{aligned}
\rho(msg_i, msg_j) = \frac{\mathbf{msg_i} \cdot \mathbf{msg_j}}{\left \lVert \mathbf{msg_i} \right \rVert \left \lVert \mathbf{msg_j} \right \rVert}
\end{aligned}
\label{eq:cosine_distance}
\end{equation}

\noindent where $\mathbf{msg_i}$ and $\mathbf{msg_j}$ denote textual embeddings, $\mathbf{msg_i} \cdot \mathbf{msg_j}$ refers to the dot product of the textual embeddings, and $\left \lVert \mathbf{msg_i} \right \rVert$ and $\left \lVert \mathbf{msg_j} \right \rVert$ denote the corresponding Euclidean norms.

\textbf{Topic Similarity:} $T(msg_x)$ denotes the singular topic assigned to message $msg_x$. As each message is only associated with one topic, a single topic is related to multiple messages. We can obtain the topic embedding $\mathbf{T_x}$ by aggregating related message embeddings. The similarity of messages with topic labels is quantified using cosine distance:

\begin{equation}
\begin{aligned}
\rho(T(msg_i), T(msg_j)) =\frac{\mathbf{T_i} \cdot \mathbf{T_j}}{\left \lVert \mathbf{T_i} \right \rVert \left \lVert \mathbf{T_j} \right \rVert},
\end{aligned}
\label{eq:topic_distance}
\end{equation}

\textbf{Unified Distance Metric:} We introduce a unified distance metric to encapsulate both textual and topical similarity:

\begin{equation}
\begin{aligned}
d(msg_i, msg_j) = \alpha \times \rho(msg_i, msg_j)\\ + (1 - \alpha) \times \rho(T(msg_i), T(msg_j)),
\end{aligned}
\label{eq:combined_distance}
\end{equation}

Here, $\alpha$ balances the influence of the textual and topical embeddings in clustering. We set $\alpha$ as 0.5 to treat both of them equally. 

\textbf{Clustering Algorithm:} We use k-means to partition messages into \(K\) clusters, optimized by a combined distance metric \(d(msg_i, msg_j)\). The centroids $\mu_k$ are re-calibrated by computing the mean of all messages designated to cluster $C_k$:

\begin{equation}
\mu_k = \frac{1}{|C_k|} \sum_{msg_i \in C_k} msg_i,
\end{equation}

This clustering prepares the ground for the ensuing $Y_-\&Y_+$ analysis aimed at neutralizing sentiment imbalance.

\subsubsection{Dominant Cluster Identification Algorithm (DCIA)}

DCIA is designed to establish a $Y_-\&Y_+$ neutralization information by focusing on critical clusters. It considers clusters' size and entropy to prioritize larger, stable ones. A pivotal aspect of the DCIA is the Memory Mechanism, which adapts strategies based on the feedback from UA garnered from their interactions. If the user agent consistently rejects the recommendation, DCIA adjusts the focus of the $Y_-\&Y_+$ neutralization to another cluster.

\textbf{Entropy Calculation for Each Cluster:} Entropy quantifies diversity in a cluster, focusing on topics and sentiments. For a given cluster \(C_k\) and topic \(\tau_i\), the average sentiment score $P'$ is:

\begin{equation}
P'(\tau_i, C_k) = \frac{\sum_{j \in C_k} o^j_{\tau_{i}}}{\sum_{j \in C_k}}
\end{equation}

\noindent
where $o^j_{\tau_{i}}$ denotes the sentiment score of message $msg_j$ towards topic $\tau_i$. When $o^j_{\tau_{i}}\in[0, 1]$, it follows that the range of $P'(\tau_i, C_k)$ will also fall within the $[0, 1]$ interval.

Now, we can compute the entropy for each cluster, considering the average sentiment scores. The entropy of cluster $C_k$ is defined as:

\begin{equation}
\label{eq:cluster_entropy}
H(C_k) = - \sum_{\tau_i \in T} P'(\tau_i, C_k) \log_2 P'(\tau_i, C_k)
\end{equation}

In Equation \ref{eq:cluster_entropy}, $H(C_k)$ measures the diversity of sentiment associated with the topics within the cluster. A higher value signifies a more diverse array of viewpoints within the cluster.

\noindent\textbf{Cluster Importance Evaluation Algorithm} We define cluster importance (\(Imp(C_k)\)) using its size (\(S_k\)) and entropy (\(H(C_k)\)). Normalized size and entropy are \(S'_k\) and \(H'(C_k)\), respectively:

\begin{equation}
S'_k = \frac{S_k}{\sum_{j=1}^n S_j}
\end{equation}

\noindent where \(n\) is the total number of clusters. Furthermore, the entropy values are normalized as follows:

\begin{equation}
H'(C_k) = \frac{H(C_k)}{\max_{j=1,\dots,n} H(C_j)}
\end{equation}

Finally, \(Imp(C_k)\) is defined as:

\begin{equation}
\textit{Imp}(C_k) = \frac{2 \cdot (S'_k \times H'(C_k))}{(S'_k + H'(C_k))}
\end{equation}

\noindent\textbf{Memory Mechanism}\label{sec:memory} The Memory Mechanism guides the $Y_-\&Y_+$ neutralization efforts toward topics within a specific cluster, informed by UA's feedback. A threshold, denoted by $\textbf{R}$, is set to limit the allowable number of rejections for messages produced by INA within the final recommendation list. If the number of rejections by a user \(v_i\) surpasses \(\textbf{R}\), INA adjusts the $Y_-\&Y_+$ neutralization target to the subsequent cluster of importance.

\subsubsection{Yin-Yang Neutralization Control Method}

The primary task of the Yin-Yang Neutralization Control (YYNC) Method is to evaluate the $Y_-\&Y_+$ balance of topics within a specified cluster and to apply balancing measures to those topics that are deemed imbalanced.

Defined in Section \ref{sec: YYND}, the $Y_-\&Y_+$ neutralization process is initiated when a target sentiment list $O_{\tau_{i}}$ = \{$o^1_{\tau_{i}}$,...,$o^m_{\tau_{i}}$\} fails to form $Y_-\&Y_+$ pairs, indicating an imbalance. For $Y_-\&Y_+$ neutralization, each sentiment in $O_{\tau_{i}}$ must pair with another (complementary sentiment). In cases of imbalance, the YYNC method seeks the most appropriate sentiment score from the database to create $Y_-\&Y_+$ pairs, as detailed in Algorithm 1, thus achieving neutralization to the sentiment list.

\begin{algorithm}[]
    \caption{Yin-Yang Neutralization Control Method.}\label{YY_algorithm}
    \begin{algorithmic}[1]
    \State Input:
    \State - Sentiment data pool for topic $\tau_{i}$: $O^{pool}_{\tau_{i}}$ = \{$o^i_{\tau_{i}}$, ..., $o^n_{\tau_{i}}$\}
    \State - $Y_-\&Y_+$ neutralization target for topic $\tau_{i}$: $O_{\tau_{i}}$ = \{$o^j_{\tau_{i}}$, ..., $o^m_{\tau_{i}}$\}
    \State - tolerant: $tol$
    \State Output: A complementary list for $Y_-\&Y_+$ neutralization: $complement\_list$ 

    \Statex

    \Function{IfBalance}{$O_{\tau_{i}}$}
         \State Sort $O_{\tau_{i}}$ by closeness to neutral sentiment 0.5
        \State Initialize $remain\_list$ to track unpaired sentiments
        \While{$\text{len}(O_{\tau_{i}}) \neq 0$}
            \State $min, max \gets \text{get\_radiation\_tolerance}(o^m_{\tau_{i}})$ 
            \Comment{The distance between ($o^m_{\tau_{i}}$, $min$) and ($o^m_{\tau_{i}}$, $max$) is the tolerance interval for $Y_-\&Y_+$ neutralization}
            \State $o^{m^+}_{\tau_{i}} \gets \text{find\_complement\_num}(O^{t}_{\tau_{i}}, min, max)$
            \If{$o^{m^+}_{\tau_{i}}$ is not None}
                \State $O_{\tau_{i}}.\text{remove}(o^{m^+}_{\tau_{i}})$
            \Else
                \State $remain\_list.\text{append}(o^m_{\tau_{i}})$
            \EndIf
        \EndWhile
    \EndFunction

    \Statex
    \Function{FindMatchScore}{$remain\_list, O^{pool}_{\tau_{i}}$}
            \State Initialize $complement\_list$ to include sentiments added for balance
            \For{$o^{t}_{\tau_{i}}$ \textbf{in} $remain\_list$}
                \State $min, max \gets \text{get\_radiation\_tolerance}(o^{t}_{\tau_{i}})$
                \State $o^{t^+}_{\tau_{i}} \gets \text{find\_complement\_num}(O^{pool}_{\tau_{i}}, min, max)$
                \If{$o^{t^+}_{\tau_{i}} \neq \text{None}$}
                    \State $complement\_list.\text{append}(o^{t^+}_{\tau_{i}})$
                    \State \Return $complement\_list$
                \Else
                    \State $min = min + tol$; $max = max + tol$
                    \State Repeat lines 23-26.
                \EndIf
            \EndFor
    \EndFunction
\end{algorithmic}
\end{algorithm}

\subsubsection{Searching}

The final phase of the $Y_-\&Y_+$ neutralization process, referred to as ``Searching'', revolves around treating the sentiment list ($complement\_list$) generated by the YYNC method as a search query. This query is subsequently utilized to retrieve relevant messages from the information message base. These retrieved messages are then finally merged with the recommendation ($M^t_{v_i}$) provided by OPA, culminating in a unified message set that is presented as a recommendation to UA.

\begin{figure}[!t]
\centering
\includegraphics[width=3.6in]{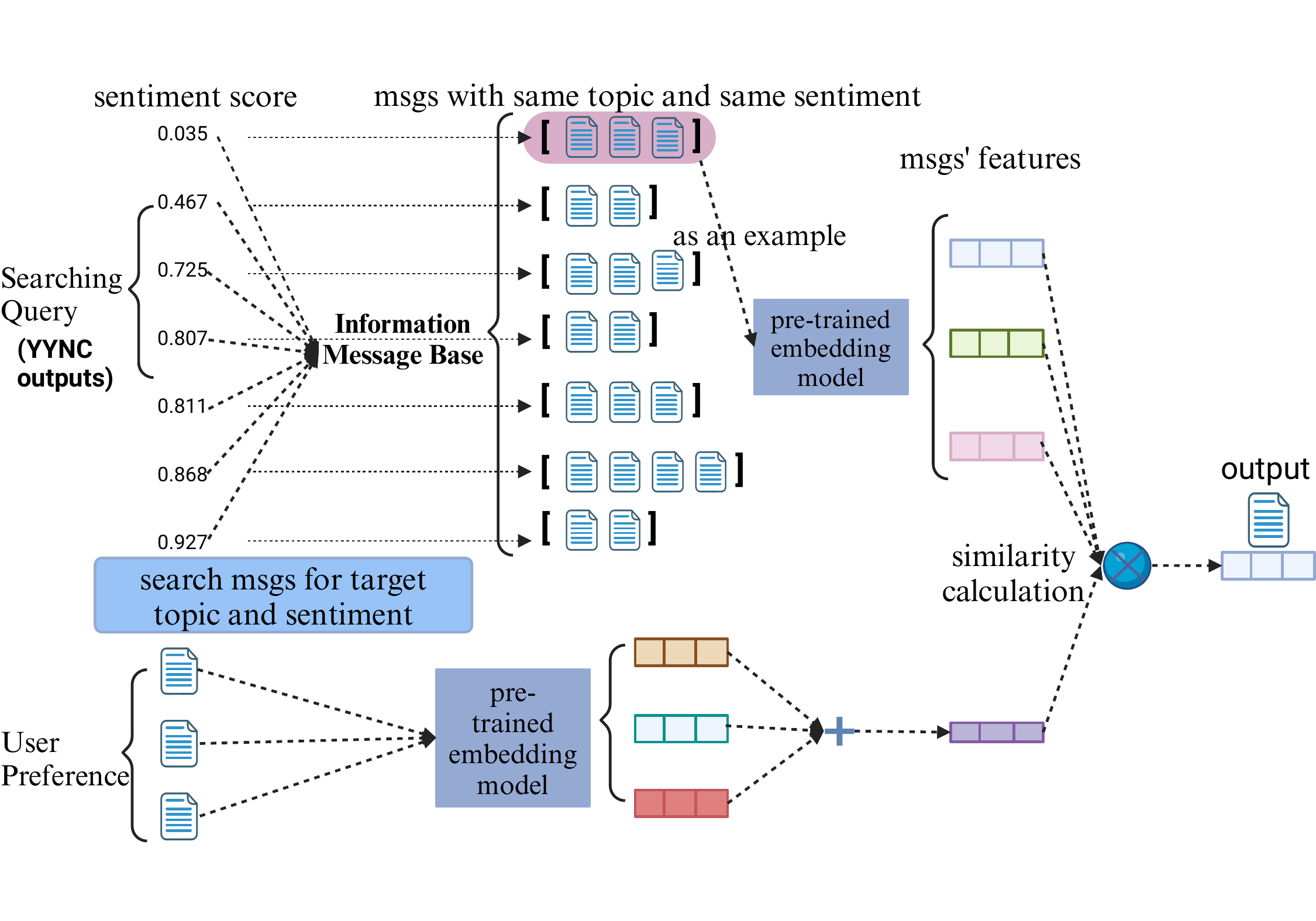}
\caption{The searching process in INA.}
\label{fig:searching}
\end{figure}

Illustrated in Figure \ref{fig:searching}, the outputs from the YYNC method function as the query for the ``Searching'' step. This query is input into the Information Message Base, initiating a search within the base to extract messages with matching topics and corresponding sentiment scores. In the depicted example of Figure \ref{fig:searching}, assuming the sentiment query value associated with topic $\tau_i$ is 0.035, we search all messages of topic $\tau_i$ with 0.035 sentiment score from the information message base. Subsequently, a pre-trained embedding model is engaged to extract features from each message. A similarity assessment is executed between the feature of each message and user preferences $\mathbf{v_i}^t$ to ensure the output message exhibits the highest similarity with user preferences. In essence, for each sentiment query, a corresponding message output is obtained. The accumulation of these outputs combined with $M^t_{v_i}$ from OPA is INA's final output, donated as $M^t_{{v_i}R}$.

\subsection{User Agent}
\label{sec:feedback}
The User Agent (UA) receives a $Y_-\&Y_+$ neutralization recommendation list from INA. It responds to this list based on user preferences from historical behaviors and decides whether to accept or reject the messages from INA. To simulate the user's responses to INA's recommendations, an offline Acceptance Probability Algorithm is employed. This algorithm assesses the similarity between preferences of UA and the messages in $M^t_{{v_i} R}$ using the cosine similarity formula:

\begin{equation}
\begin{aligned}
\rho(v_i^t, M^t_{{v_i} R}) = \frac{\mathbf{v_i}^t \cdot \mathbf{M}^t_{{v_i} R}}{||\mathbf{v_i}^t|| ||\mathbf{M}^t_{{v_i} R}||},
\end{aligned}
\label{eq:cosine_distance_UF}
\end{equation}

\noindent 
where $\mathbf{M}^t_{{v_i} R}$ represents the integrated embedding of all messages in $M^t_{{v_i} R}$, and $\mathbf{v_i}^t$ signifies $v_i$'s preferences at time $t$.

This algorithm incorporates a random value to model $v_i$'s responses. If the cosine similarity $\rho(v_i^t, M^t_{{v_i} R})$ exceeds this random threshold, it indicates that $v_i$ accepts the recommended message. Otherwise, the message is rejected. Based on these acceptance or rejection decisions, feedback from UA is forwarded to OPA for the generation of next-round messages. Simultaneously, this feedback updates the user preference repository, enhancing the accuracy of future recommendations from INA.

\section{Experiments and Analysis}\label{sec:experiment}

In our research, we aim to thoroughly assess the AbIN model's efficacy in fostering sentiment diversity, realizing $Y_-\&Y_+$ neutralization, and consequently mitigating filter bubbles. To this end, we carried out three interconnected experiments. These experiments investigate AbIN's performance on recommendation diversity, precision, and $Y_-\&Y_+$ neutralization.

\subsection{Datasets and Settings}

We utilize two real-world datasets, i.e., the Microsoft News dataset (MIND) \footnote{https://msnews.github.io/} and IMDB \footnote{https://www.kaggle.com/datasets/meastanmay/imdb-dataset?select=tmdb\_5000\_movies.csv/} dataset. MIND is a public news recommendation dataset, encompassing user interaction data gathered from Microsoft News. MIND comprises data from 5,000 users, encompassing 230,117 user reading records and 51287 news. IMDB is a movie recommendation dataset. We use the rating records provided by the dataset. It comprises 25,000 movie rating records from 333 users and a range of 2586 movies.

\textbf{Settings:} Considering the inherent impracticality and substantial cost of conducting online testing for researchers, we have developed an offline simulation approach to evaluate the proposed method. The Acceptance Probability Algorithm described in Section \ref{sec:feedback} simulates user feedback for recommendations.

\subsection{Evaluation Metrics}
We assess the AbIN system using three key metrics: Diversity, Accuracy, and Yin-Yang Neutralization Degree (best\_diff).

Diversity: Measures the range of sentiments across recommendations, indicating a system's ability to offer varied perspectives. It consists of sentiment coverage and repetition rate (RR), where coverage is the sum of sentiments for a particular topic in the recommendation list divided by the sum of sentiments for that topic in the data pool and then averaged. The RR calculates the average sentiment redundancy for a specific topic, as the topic is $\tau_i$, $RR = \frac{\sum_{i=1}^n RR_{\tau_i}}{n}$.

Accuracy: Reflects how recommendations align with user preferences. It's evaluated through Hit (1 if at least one message is accepted in a recommendation, 0 otherwise) and Precision (Pre), which is the ratio of accepted items in the recommendation.
     
best\_diff: Quantifies the degree of $Y_-\&Y_+$ neutralization in recommendation systems, we introduce a metric defined by the formula: $\left| \sum_{x=1}^{n}f(o^x_{\tau_{i}})-\sum_{y=1}^{m}o^y_{\tau_{i}}\right|$, where $f$ represents the curve function detailed in Equation \ref{eq:stype}. Here, $o^x_{\tau_{i}}$ and $o^y_{\tau_{i}}$ denote the individual sentiments within the list $O_{\tau_{i}}$, corresponding to $Y_-$ and $Y_+$ respectively. The variables $m$ and $n$ are the counts of $Y_-$ and $Y_+$ scores in the list. A ``best\_diff'' value of 0 signifies that the list $O_{\tau_{i}}$ has achieved the perfect neutralization.

\subsection{Baselines}
We assess the performance of AbIN in comparison with several established baseline methods:

\begin{itemize}
    \item NGCF \cite{wang2019neural}: Enhances recommendation by modeling collaborative signals with user-item graphs.
    \item LGCN \cite{he2020lightgcn}: Focuses on neighborhood aggregation for efficient training; outperforms NGCF.
    \item DGCF \cite{wang2020disentangled}: Disentangles user-item relationships to capture user intent diversity.
    \item ENMF \cite{chen2020efficient}: Employs three new optimization methods in a neural matrix factorization framework for better performance.
    \item SGL \cite{wu2021self}: Incorporates self-supervised learning into traditional GCNs for improved accuracy and robustness.
\end{itemize}

For each baseline, we also explore its integration with the AbIN model, denoted as ``Method$_{AbIN}$'' (e.g., SGL$_{AbIN}$), resulting in 10 comparative baselines to assess the impact of AbIN on overcoming filter bubbles and achieving neutralization goals.

\subsection{Experiment 1: Evaluation of Diversity and Accuracy}

The first experiment evaluates the AbIN model's impact on enhancing diversity and accuracy within preference-based recommendation systems. By leveraging metrics such as Coverage and RR as metrics for diversity, and Hit and Pre for accuracy, our objective is to show how the AbIN model enhances sentiment diversity without significantly compromising recommendation precision. The experiment conducted on the MIND and IMDB datasets involved 10 recommendation models applied to 5 randomly selected users from each dataset (U45, U46, U103, U10022, U23393 from MIND and U22, U62, U63, U88, U109 from IMDB), illustrating the benefits of incorporating the AbIN model in overcoming filter bubbles and fostering a neutral recommendation environment. The average evaluation results are shown in Table \ref{tab:all_users}, where ``$\Delta$(\%)'' denotes the percentage change due to AbIN.

\begin{table*}[!t]
\renewcommand{\arraystretch}{0.5}
\caption{Comparative Analysis of Recommendation Models with and without the AbIN Enhancement \label{tab:all_users}}
\centering
\resizebox{1\columnwidth}{!}{%
\begin{tabular}{l|cccc|cccc}
\toprule\midrule
\multicolumn{1}{c|}{\multirow{10}{*}{Models}} & \multicolumn{4}{c|}{\textbf{MIND}} & \multicolumn{4}{c}{\textbf{IMDB}} \\ 
\cmidrule(lr){2-5} \cmidrule(lr){6-9}
\multicolumn{1}{c|}{} & \multicolumn{2}{c}{\textbf{Diversity}} & \multicolumn{2}{c|}{\textbf{Accuracy}} & \multicolumn{2}{c}{\textbf{Diversity}} & \multicolumn{2}{c}{\textbf{Accuracy}} \\ \cmidrule(lr){2-5} \cmidrule(lr){6-9}

\multicolumn{1}{c|}{} & \textbf{Coverage} & \textbf{RR} & \textbf{Pre} & \textbf{Hit} & \textbf{Coverage} & \textbf{RR} & \textbf{Pre} & \textbf{Hit} \\ \midrule\midrule

SGL & 0.0009 & 0.0349 & 0.4920 & 1 & 0.0136 & 0.0031 & 0.4619 & 1\\
SGL$_{AbIN}$ & 0.0012 & 0.0269 & 0.5012 & 1 & 0.0145 & 0.0015 &  0.4902 & 1\\\midrule
$\Delta$(\%) & $\uparrow$21.67 &  $\downarrow$ 29.56 & $\uparrow$ 1.84 & - & $\uparrow$5.708& $\downarrow$ 109.52 & $\uparrow$ 5.77 & - \\\midrule\midrule

NGCF & 0.0011 & 0.0399  & 0.4660 & 1 & 0.0135 & 0.0030 & 0.4880 & 1\\
NGCF$_{AbIN}$ & 0.0014 & 0.0315 & 0.4454 & 1 &  0.0193 &  0.0025 &  0.5006 & 1\\\midrule
$\Delta$(\%)& $\uparrow$ 18.06 & $\downarrow$ 21.11 & $\downarrow$ 4.63 &  - & $\uparrow$ 29.68 & $\downarrow$ 20.39 & $\uparrow$ 2.52 & - \\\midrule\midrule

LGCN & 0.0010 & 0.0413 & 0.538  & 1 & 0.0116  & 0.0122 & 0.516 & 1\\
LGCN$_{AbIN}$ & 0.0012 & 0.0252 &  0.5232  & 1 & 0.0136 & 0.0066 & 0.4844 & 1\\\midrule
$\Delta$(\%) &  $\uparrow$ 20.03 & $\downarrow$ 63.92   & $\downarrow$ 2.81 & - & $\uparrow$ 14.61 & $\downarrow$ 84.22 & $\uparrow$ 6.516 & -  \\\midrule\midrule

DGCF & 0.0012 & 0.0471 & 0.49 & 1  & 0.0143 &  0.0044 &  0.53 & 1\\
DGCF$_{AbIN}$ & 0.0014 &  0.0328 & 0.460  & 1 & 0.0152 & 0.0032 & 0.5140 & 1 \\\midrule
$\Delta$(\%) & $\uparrow$ 14.78 & $\downarrow$43.59& $\downarrow$6.52 & - & $\uparrow$5.89 & $\downarrow$37.5 &  $\downarrow$ 3.09& - \\\midrule\midrule

ENMF & 0.0010 &  0.0333 & 0.490 & 1 & 0.0138 & 0.0025  &  0.512 & 1 \\
ENMF$_{AbIN}$ & 0.0012 &  0.0259 & 0.5196 & 1 & 0.0161 & 0.0021 & 0.5165  & 1 \\\midrule
$\Delta$(\%) & $\uparrow$ 9.308 & $\downarrow$ 28.61 & $\uparrow$ 5.704 & -  & $\uparrow$ 13.85 & $\downarrow$ 18.54 & $\uparrow$ 0.875 & - \\\midrule\bottomrule
\end{tabular}
}
\end{table*}

Results indicate that incorporating the AbIN model significantly increases sentiment coverage across both datasets and reduces sentiment repetition. This leads to a richer user experience by offering a broader range of sentiments. Despite mixed outcomes in precision, the findings underscore the AbIN model's ability to balance diversity and accuracy effectively. Overall, this experiment underscores the AbIN model's role in enhancing the diversity of preference-based recommendation systems. By broadening the variety of content recommended to users, the model addresses critical challenges such as filter bubbles.

\subsection{Experiment 2: Neutralization Evaluation}

\begin{figure*}[!t]
\centering
\subfloat[]{\includegraphics[width=3.5in]{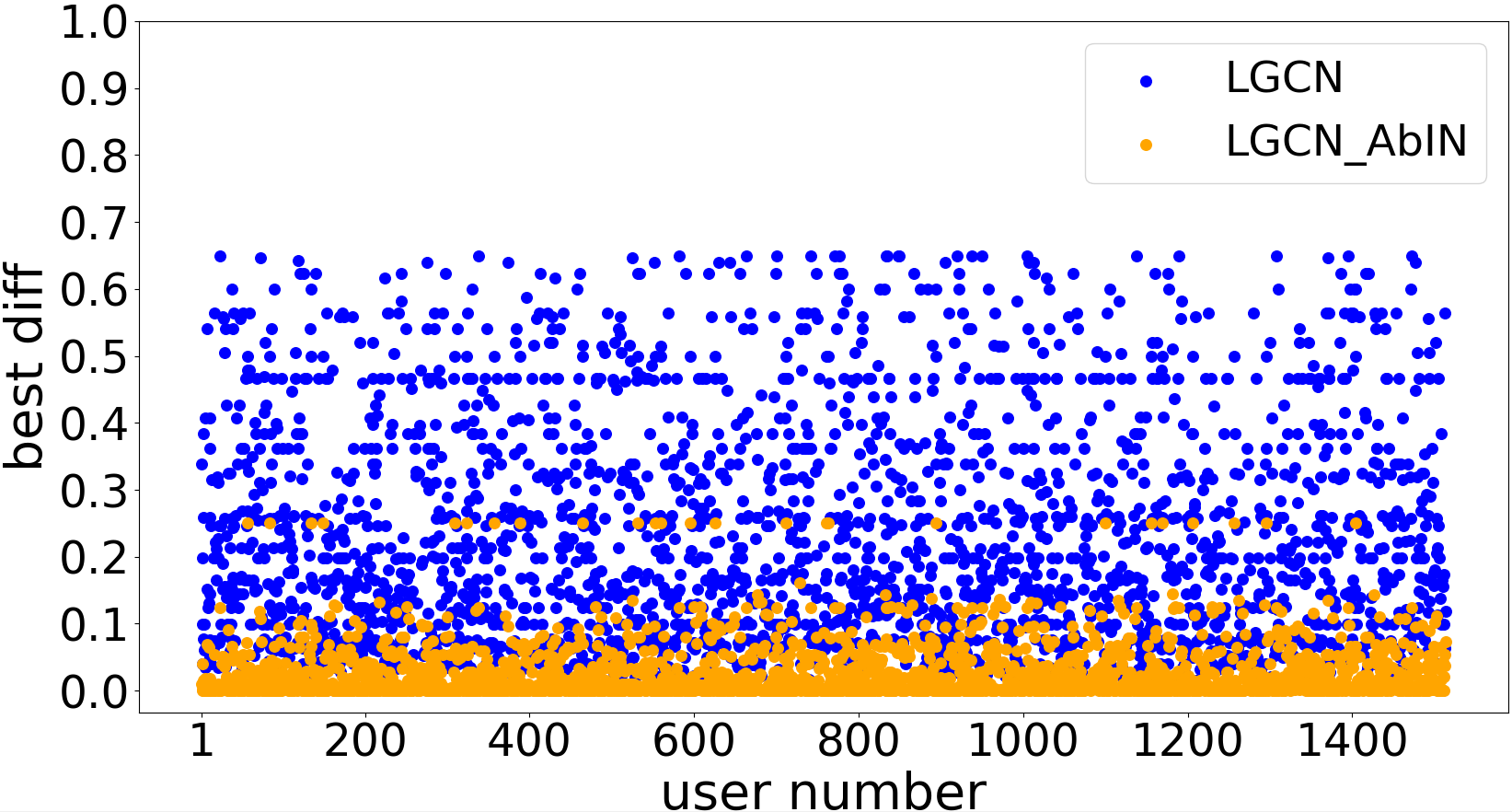}%
\label{fig:yy_MIND}}
\hfil
\subfloat[]{\includegraphics[width=3.5in]{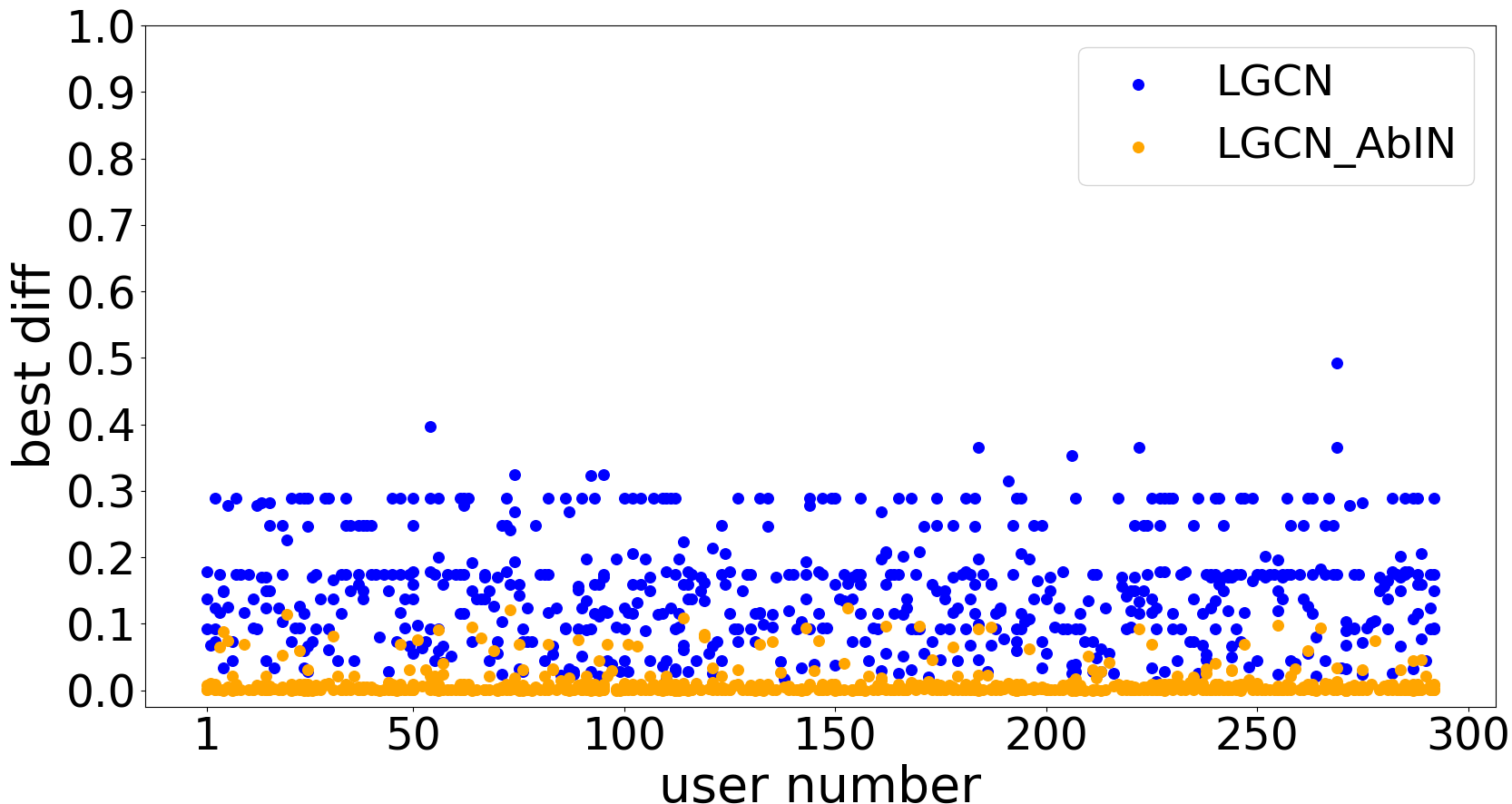}%
\label{fig:yy_IMDB}}
\caption{An analysis of the efficacy of the AbIN model in the context of a single-instance Yin-Yang neutralization task on two datasets. (a) Yin-Yang neutralization evaluation on MIND dataset. (b) Yin-Yang neutralization evaluation on IMDB dataset.}
\label{fig:neutralzation_evaluation}
\end{figure*}

This experiment evaluates the AbIN model's capability to enhance neutralization in recommendation systems, specifically aiming to balance  $Y_-$ and  $Y_+$ to provide a more neutral recommendation environment and reduce the impact of filter bubbles.  The ``best\_diff'' matrix serves as a quantitative measure of neutralization extent. This study includes two datasets: 2,000 users were randomly selected from MIND, and all users from IMDB were included.

In the scatter plots referenced (Figure \ref{fig:neutralzation_evaluation}), the x-axis indicates user numbers, and the y-axis represents the degree of $Y_-\&Y_+$ neutralization (best\_diff) across recommended topics for each user in a single recommendation. The figures reveal that the AbIN-enhanced model (LGCN$_{AbIN}$), depicted by yellow dots, significantly improves $Y_-\&Y_+$ neutralization compared to the LGCN model (blue dots), with many instances nearing perfect neutralization (approaching 0). This suggests that integrating AbIN aids in achieving a more neutral recommendation environment.

The experimental results underscore the AbIN model's impact on improving $Y_-\&Y_+$ neutralization within recommendation systems. By consistently lowering the ``best\_diff'' values across a broad user base, the AbIN-enhanced model demonstrates its superiority over preference-based approaches.

\subsection{Experiment 3: Impact of Cluster Sizes}

\begin{table}[!t]
\renewcommand{\arraystretch}{0.8}
\caption{Comparison of LGCN and LGCN$_{AbIN}$ on MIND and IMDB datasets \label{tab:comparison}}
\centering
\resizebox{1.0\columnwidth}{!}{
  \begin{tabular}{c|cccc|cccc}\hline
  \multicolumn{1}{c|}{\multirow{4}{*}{}} & \multicolumn{4}{c|}{\textbf{MIND}} & \multicolumn{4}{c}{\textbf{IMDB}} \\ \hline
    cluster & time(m) & LGCN & LGCN$_{AbIN}$ & Improve(\%) & time(m) & LGCN & LGCN$_{AbIN}$ & Improve(\%) \\
    1 & 76.58 & 0.2332 & 0.0244 & 89.54 & 0.598 & 0.7287  & 0.0165 & 97.72 \\
    \textbf{2} & 38.56 & 0.2358 & 0.0203 & 91.38 & 0.454 & 0.6622 & 0.0110 & \textbf{98.32} \\
    3 & 34.06 & 0.2384 & 0.0192 & 91.97 & 0.400 & 0.6650 & 0.0111 & 98.31 \\
    4 & 30.41 & 0.2442 & 0.0191 & 92.18 & 0.355 & 0.8391 & 0.0264 & 96.84 \\
    5 & 30.30 & 0.2333 & 0.0180 & 92.27 & 0.358 & 0.9169 & 0.0390 & 95.74 \\
    \textbf{6} & 29.18 & 0.2293 & 0.0174 & \textbf{92.32} & 0.364 & 0.9388 & 0.0431 & 95.40 \\
    7 & 28.86 & 0.2263 & 0.0230 & 89.85 & 0.370 & 0.9750 & 0.0464 & 95.23 \\
    8 & 26.34 & 0.2299 & 0.0225 & 90.20 & 0.379 & 0.9697 & 0.0503 & 94.80 \\\hline
  \end{tabular}
  }
\end{table}

In this parameter experiment, we assess the impact of varying cluster sizes on the speed and extent of $Y_-\&Y_+$ neutralization by the recommendation system, which is configured to suggest 10 messages per recommendation. This setup remains consistent with that of Experiment 2, except for the cluster sizes examined.

The table shows an acceleration in the neutralization process as the cluster size increases, as noted in the ``time'' column.  This column quantifies the extent of $Y_-\&Y_+$ neutralization (reflected by ``best\_diff'') achieved by the AbIN-enhanced model compared to the LGCN model. Despite minor fluctuations, cluster size 6 is selected for the MIND dataset due to its optimal performance. Conversely, the IMDB dataset exhibits a significant shift in neutralization performance. The ``Improve(\%)'' diminishes from 98.32\% to 94.80\% as cluster size increases and stabilizes from cluster size 4 onwards. This suggests that the messages within the targeted cluster reach a stable beyond this point. Cluster size 2 is chosen for the IMDB dataset, which demonstrates the best performance on $Y_-\&Y_+$ neutralization.

\subsection{Discussions}

Our study involves three interconnected experiments to offer an extensive evaluation of the effectiveness of the proposed AbIN model. These experiments collectively illuminate the AbIN model's multifaceted capabilities, including its performance on recommendation diversity, precision,  and $Y_-\&Y_+$ neutralization. The AbIN-based recommendation algorithms reveal their ability to mitigate filter bubbles by broadening sentiment recommendation spectrums and providing a more balanced perception. The experimental results emphasize the model's potential for practical applications in recommendation systems without altering original algorithms. However, while our findings emphasize AbIN's potential, it's essential to acknowledge potential limitations. Its applicability across real-world implementation challenges needs to be further explored. Future research could probe deeper into these aspects, ensuring AbIN's robustness in diverse online environments.

\section{Conclusion and Future Work}
\label{sec:conclusion}
This paper introduces AbIN, an innovative agent-based model inspired by the $Y_-\&Y_+$ theory, that aims to mitigate filter bubbles in existing recommendation systems. Distinct from prevalent solutions that modify recommendation algorithms, AbIN incorporates multiple agents into the recommendation task. Each agent within the AbIN model is designed to be both independent and interconnected. By $Y_-\&Y_+$ analysis and neutralization, the model counteracts imbalanced information perception, fostering a more balanced sentiment perspective. Experiments showcased AbIN’s efficacy in promoting sentiment diversity, achieving $Y_-\&Y_+$ neutralization, and thus alleviating filter bubbles. To conclude, AbIN offers a novel solution to mitigate filter bubbles generated by preference-based recommendation systems. Future research avenues should delve into refining its precision, and real-world implementation, exploring its dynamic adaptations, and assessing the impact on users.

\section*{Acknowledgments}
The authors would like to acknowledge the financial support from Callaghan Innovation (CSITR1901, 2021), New Zealand, without which this research would not have been possible. We are grateful for their contributions to the advancement of science and technology in New Zealand. The authors would also like to thank CAITO.ai for their invaluable partnership and their contributions to the project.

\newpage

\bibliographystyle{IEEEtran}
\bibliography{ref}

\end{document}